\documentclass[
aps,
prl,
twocolumn,
superscriptaddress,
floatfix
longbibliography
]{revtex4-1}
\usepackage[dvipdfmx]{graphicx}
\usepackage{verbatim}
\usepackage{epsf}
\usepackage{natbib}
\usepackage{dcolumn}
\usepackage{bm}
\usepackage{color} 
\usepackage{ulem}
\usepackage{amsmath}
\usepackage{amssymb}
\usepackage{txfonts}
\usepackage[
colorlinks=true,
citecolor=blue,
urlcolor=blue,
setpagesize=false]{hyperref}

\hypersetup{
  colorlinks=true,
  linkcolor=[rgb]{0.60,0.00,0.0},
  citecolor=[rgb]{0.0,0.0,0.6},
  urlcolor=[rgb]{0.0,0.0,0.6},
  setpagesize=false
}
\begin{document}

\title{
  Correlation-Driven
  Dimerization and  Topological Gap Opening 
  in Isotropically Strained Graphene
}

\author{Sandro~Sorella}
\affiliation{International School for Advanced Studies (SISSA), Via Bonomea 265, 34136, Trieste, Italy} 
\affiliation{Democritos Simulation Center CNR--IOM Istituto Officina dei Materiali, Via Bonomea 265, 34136 Trieste, Italy}            
\affiliation{Computational Materials Science Research Team, RIKEN Center for Computational Science (R-CCS),  Hyogo 650-0047,  Japan}

\author{Kazuhiro~Seki}
\affiliation{International School for Advanced Studies (SISSA), Via Bonomea 265, 34136, Trieste, Italy} 
\affiliation{Computational Materials Science Research Team, RIKEN Center for Computational Science (R-CCS),  Hyogo 650-0047,  Japan}
\affiliation{Computational Condensed Matter Physics Laboratory, RIKEN Cluster for Pioneering Research (CPR), Saitama 351-0198, Japan}

\author{Oleg O. Brovko}
\affiliation{The Abdus Salam International Centre for Theoretical Physics (ICTP), Strada Costiera 11, 34151, Trieste, Italy} 

\author{Tomonori~Shirakawa}
\affiliation{International School for Advanced Studies (SISSA), Via Bonomea 265, 34136, Trieste, Italy} 
\affiliation{Computational Materials Science Research Team, RIKEN Center for Computational Science (R-CCS),  Hyogo 650-0047,  Japan}
\affiliation{Computational Condensed Matter Physics Laboratory, RIKEN Cluster for Pioneering Research (CPR), Saitama 351-0198, Japan}
\affiliation{Computational Quantum Matter Research Team, RIKEN, Center for Emergent Matter Science (CEMS), Saitama 351-0198, Japan}

\author{Shohei~Miyakoshi}
\affiliation{Computational Quantum Matter Research Team, RIKEN, Center for Emergent Matter Science (CEMS), Saitama 351-0198, Japan}

\author{Seiji~Yunoki}
\affiliation{Computational Materials Science Research Team, RIKEN Center for Computational Science (R-CCS),  Hyogo 650-0047,  Japan}
\affiliation{Computational Condensed Matter Physics Laboratory, RIKEN Cluster for Pioneering Research (CPR), Saitama 351-0198, Japan}
\affiliation{Computational Quantum Matter Research Team, RIKEN, Center for Emergent Matter Science (CEMS), Saitama 351-0198, Japan}

\author{Erio Tosatti}            
\affiliation{International School for Advanced Studies (SISSA), Via Bonomea 265, 34136, Trieste, Italy} 
\affiliation{Democritos Simulation Center CNR--IOM Istituto Officina dei Materiali, Via Bonomea 265, 34136 Trieste, Italy}            
\affiliation{The Abdus Salam International Centre for Theoretical Physics (ICTP), Strada Costiera 11, 34151, Trieste, Italy}

\begin{abstract}
The phase diagram of isotropically expanded graphene cannot be correctly predicted by ignoring either electron correlations, or mobile carbons, or the effect of applied stress,  as was done so far. We calculate the ground state enthalpy (not just energy) of strained graphene by an accurate off-lattice quantum Monte Carlo (QMC) correlated ansatz of great variational flexibility. Following undistorted semimetallic graphene at low strain, multideterminant Heitler-London correlations stabilize between $\simeq$8.5\% and $\simeq$15\% strain an insulating Kekul\'e-like dimerized (DIM) state. Closer to a crystallized resonating-valence bond than to a Peierls state, the DIM state prevails over the competing antiferromagnetic insulating (AFI) state favored by  density-functional calculations which we conduct in parallel.  The DIM stressed graphene insulator, whose gap is predicted to  grow in excess of 1 eV before failure near 15\% strain,  is topological in nature, implying under certain conditions 1D metallic interface states lying in the bulk energy gap. 
\end{abstract}

\date{\today}

\maketitle

\begin{figure}
  \begin{center}
    \includegraphics[width=.8\columnwidth]{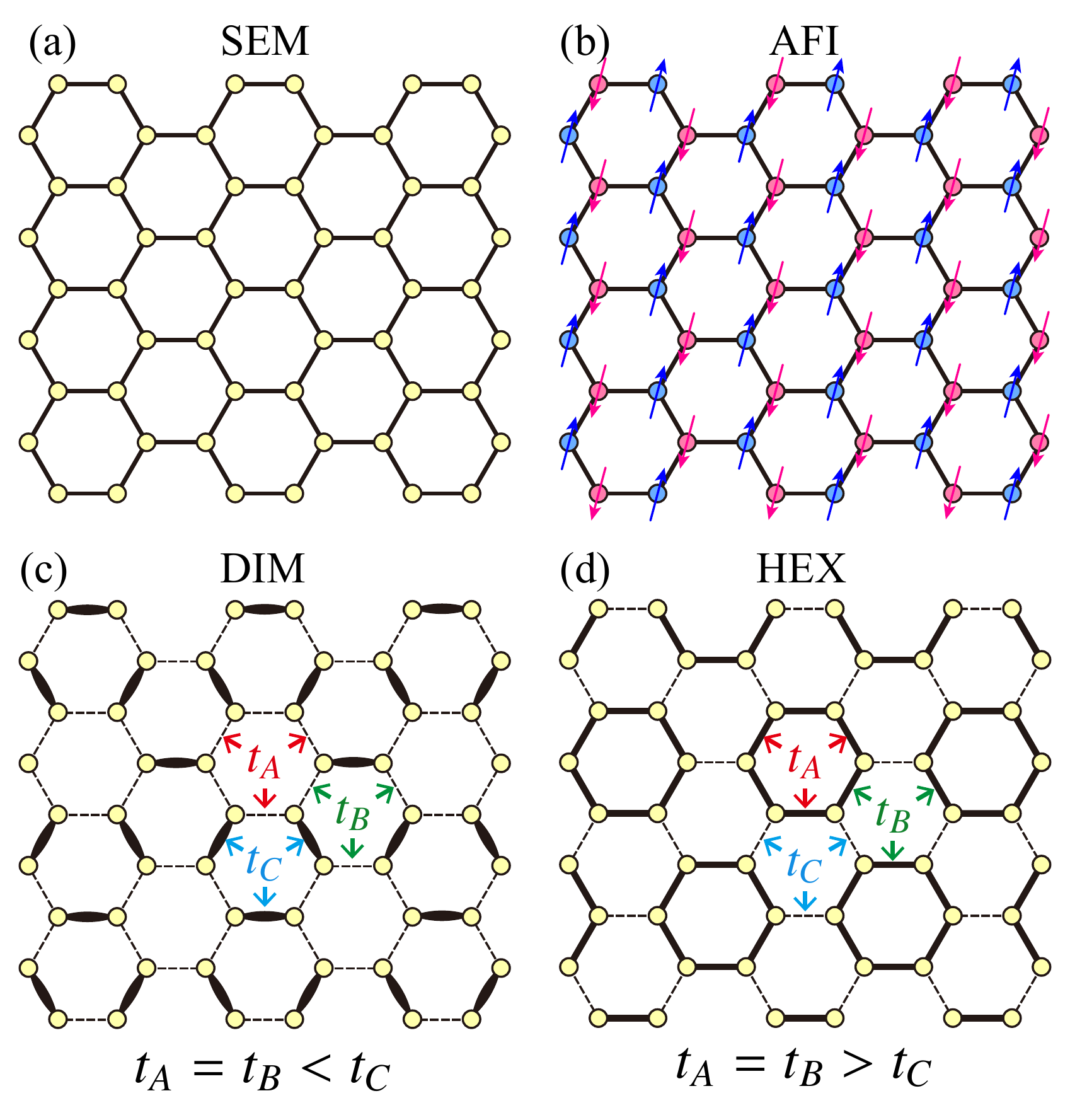}
    \caption{
      \label{fig1}
      (a) SEM honeycomb, semimetallic; 
      (b) AFI honeycomb antiferromagnetic insulator; 
      (c) DIM dimerized Kekul\'e-like insulator; 
      (d) HEX distorted hexagonal insulator. 
      There are two carbons per unit cell in (a) and (b), six in (c) and (d). 
      Following Ref.~\cite{Frank2011}, $t_A$, $t_B$, and $t_C$ schematically denote hopping integrals magnitudes.
    }
  \end{center}
\end{figure}

In graphene, which current technology 
strives to employ in electronics,  
an insulating state does not naturally occur.
Strain engineering has long been considered as providing
mechanisms to pry open the symmetry-induced zero gap 
of the original semimetallic graphene [SEM, see Fig.~\ref{fig1}(a)] honeycomb structure~\cite{newreview}. 
Among them, a nonisotropic three-directional strain was suggested~\cite{guinea} 
and verified~\cite{newreview, Jiang2017, Naumis2017,prb1,ref63} to introduce a gauge field and a gap.

An insulating state could alternatively be achieved in graphene by simple isotropic expansive strain.
Experimentally, indentation experiments suggested that graphene can be isotropically stressed until mechanical failure near $22.5\%$ strain, corresponding to a tensile stress around 50 N/m~\cite{Lee2008}. 
No evidence of structural or electronic transition occurring during expansion was provided. 
Theoretically, idealized  rigid-honeycomb Hubbard model,  quantum Monte Carlo (QMC) calculations had long 
suggested~\cite{Sorella_EPL1992, Meng2010,Chang_PRL2012, Sorella_SR2012,Otsuka_PRX2016} 
band narrowing and increased effective electron-electron repulsion could push the correlated $\pi$-electron system 
towards an undistorted honeycomb antiferromagnetic insulator [AFI, see Fig.~\ref{fig1}(b)]. 
Spin-polarized density-functional-theory (DFT) calculations~\cite {Lee2012, newreview} 
as well as  
rigid-lattice QMC simulations~\cite{Tang2015,ChenWagner2017} indeed suggest 
a SEM-AFI  crossing of total energies with a semimetal-insulator transition around $8-10\%$ strain.  
Alternatively,  isotropically stressed graphene could distort to form Peierls or  Kekul\'e-like 
dimerized [DIM, see Fig.~\ref{fig1}(c)] states, discussed by detailed DFT phonon calculations~\cite{Lazzeri2008,Marianetti} and by symmetry considerations~\cite{Frank2011}, with a unit-cell size increase from two to six carbons, and an electronic gap proportional to the dimerization magnitude. 
The DIM distortion scenario is nevertheless  denied by spin-polarized DFT calculations where the AFI state has lower energy than DIM. 

All this work thus leaves the electronic and structural phase diagram of 
isotropically strained graphene in a state of uncertainty, on two separate accounts. 
First, the strong band narrowing and increased role of strong electron correlations, 
improperly treated by DFT,  calls for a novel  QMC description capable of describing 
real strained and deformable graphene, a goal never attained so far.    
Second, the phase diagram under stress must be obtained by comparing enthalpies,  
therefore including  the stress-strain term, rather than just total energies, 
as was universally done so far.  Because the stress-strain equation of state is different 
for different phases, the correct phase diagram will not be 
identical to that suggested by minimizing total energy alone. 
Here we implement accurate QMC enthalpy calculations,  reaching 
a  highly instructive phase diagram for isotropically strained graphene, 
that is found to differ from that predicted by the best, spin-polarized, DFT.

Main QMC calculations were conducted based on 
a variational wavefunction (JAGP),  
known to be accurate and reliable in the description of strong electron correlations  
from  small molecules~\cite{casulaagp} 
to realistic crystalline systems~\cite{note_zenwater}, 
\begin{equation}
  \Psi_{{\rm JAGP}}= {\cal J} ({\bf r}_1, {\bf r}_2, \cdots, {\bf r}_N) 
  \Psi_{\rm AGP}({\bf r}_1 \sigma_1, {\bf r}_2 \sigma_2 , \cdots, {\bf r}_N \sigma_N ), 
  \label{wf}
\end{equation}
where ${\bf r}_i$ and $\sigma_i$, for $i=1,\cdots,N$, are the spatial and the spin coordinates of the electrons. Here  
${\cal J}= \prod\limits_{i<j} \exp \left[ u({\bf r}_i,{\bf r}_j) \right]$
is the Jastrow factor, symmetric under all particle permutations, while  the determinantal part is 
the antisymmetrized geminal power (AGP): 
$
  \Psi_{\rm AGP}={\cal A}
  f( {\bf r}_1,{\bf r}_2) \chi_{\sigma_1,\sigma_2}
  \cdots
  f( {\bf r}_{N-1},{\bf r}_N) \chi_{\sigma_{N-1},\sigma_N},
$
where ${\cal A}$ is the antisymmetrizer and the  product $f({\bf r},{\bf r}')\chi(\sigma,\sigma^\prime)$ 
describes a singlet valence-bond electron pair with an orbital-symmetric function $f({\bf r},{\bf r}')$ and  
a spin-antisymmetric one 
$\chi_{\sigma,\sigma^\prime} 
= {1 \over \sqrt{2}} (\delta_{\sigma,\uparrow} \delta_{\sigma^\prime,\downarrow}- \delta_{\sigma,\downarrow} \delta_{\sigma^\prime,\uparrow})$.  
$\Psi_{\rm AGP}$  reduces to a Slater determinant  with a particular choice of the pairing function~\cite{casularvb},  implying therefore a description of  the electron correlation better  than 
those based on any Jastrow-Slater ansatz~\cite{rmpmitas}.  
The variational freedom contained in the  $\Psi_{\rm JAGP}$ ground state naturally
permits a quantitative distinction  between the spin and charge correlations~\cite{capello}. 
Parallel reference 
DFT calculations were also performed with HSE6 
exchange-correlation functional, projector augmented-wave treatment of core levels~\cite{Blochl1994} and a plane-wave basis set~\cite{Kresse1996} as implemented in the Vienna ab-initio simulation package
(VASP)~\cite{Kresse1993,Kresse1996}, 
with energy cutoff of 600 ${\rm eV}$.  

All calculations~\cite{SM}
were conducted with 
$N_{\rm atom}=$
24 carbon atoms 
forming four six-atom unit cells of a planar deformable honeycomb lattice whose average interatomic spacing $a$ 
was successively expanded relative to the zero-stress value $a_0$.
A fully accurate $k-$point average is obtained by boundary-condition twisting.

\begin{figure}
  \begin{center}
    \includegraphics[width=.95\columnwidth]{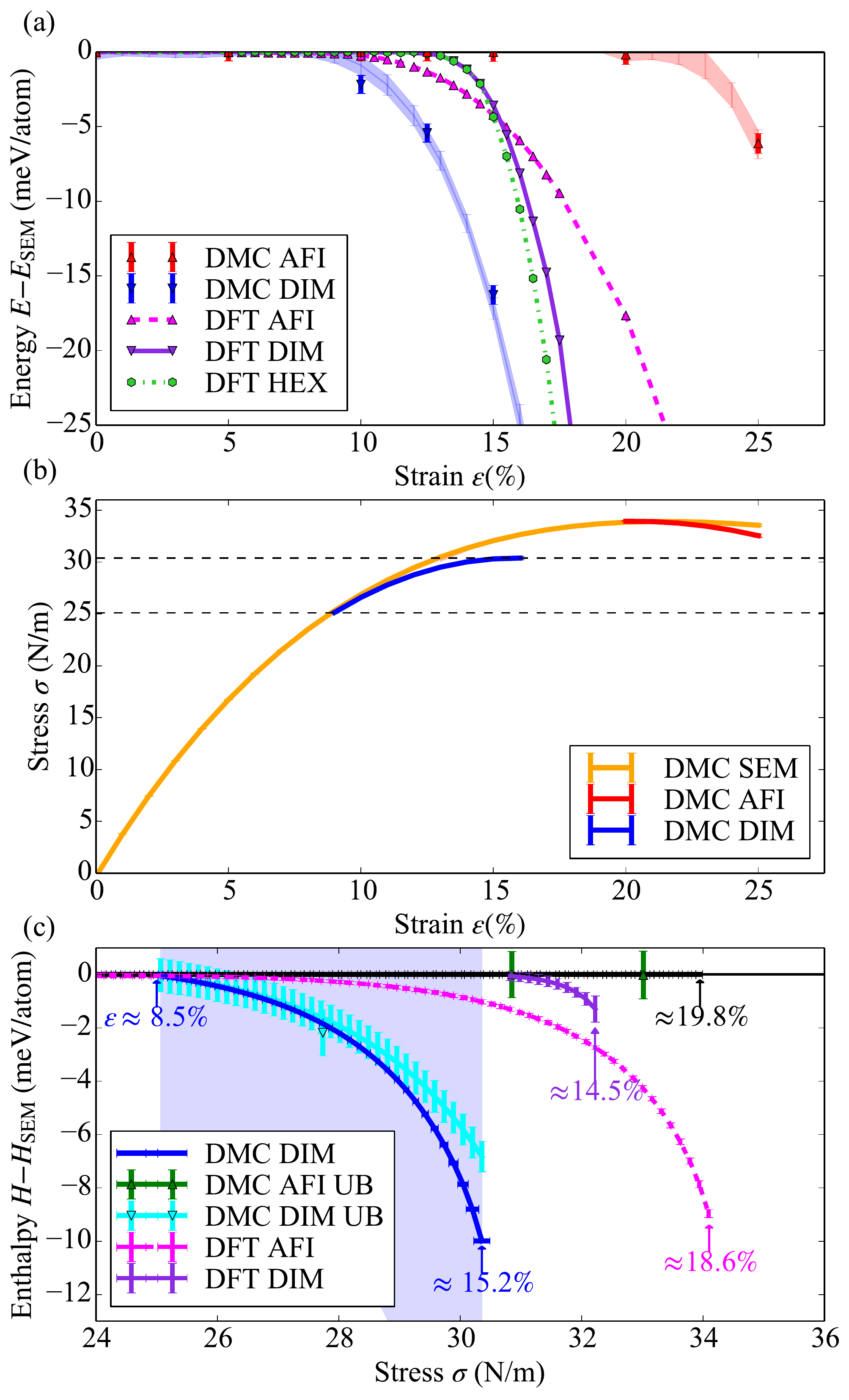}
    \caption{
      \label{fig2}
      (a) Ground state energy $E$ relative to the SEM phase $E_{\rm SEM}$  obtained 
      as a function of strain $\epsilon$ by DMC in comparison  with DFT for the DIM, AFI, (HEX) phases. 
      (b) Stress ($\sigma$)-strain ($\epsilon$) equation of state curve for strained graphene obtained by fitting DMC energies.
      Dashed lines  mark the transition stress values $\sigma_{\rm l}$ and $\sigma_{\rm u}$ for SEM-DIM (continuous).  
      (c) Enthalpy $H$ of strained graphene relative to that of the SEM phase $H_{\rm SEM}$ for increasing tensile stress $\sigma$. 
      The blue-shaded region indicates the error bars on the enthalpies for DIM and AFI phases by DMC.  
      Upper bounds of Eq.~(\ref{eq.ub}) for the DIM and AFI enthalpies  also shown (DIM UB and AFI UB) greatly reduce the error bars. 
      The corresponding strain $\epsilon$ at selected points and phases (indicated by arrows) are also shown.  
    }
  \end{center}
\end{figure}

Figure~\ref{fig2}(a) presents the  total energy gain of all ordered or distorted states 
relative to the undistorted, semimetallic, nonmagnetic SEM  state, $E-E_{\rm SEM}$, as a function of 
isotropic strain $\epsilon = (a-a_0)/a_0$, from both diffusion Monte Carlo (DMC) and DFT calculations. 
Figure~\ref{fig2}(b) shows the DMC-calculated tensile stress, yielding the 2D equations of state of expanded graphene.  
In DFT, the AFI state [Fig.~\ref{fig1}(b)] yields the lowest energy above $\epsilon \approx 7\%$, 
and represents the ground state until $\epsilon \approx 15\%$.  
Near 
15\% strain, DFT energetics predicts a Kekul\'e DIM state  [Fig.~\ref{fig1}(c)] 
to take over very briefly from AFI, just before turning itself unstable and leading  
to mechanical failure, in agreement with earlier DFT phonon calculations~\cite{Marianetti}. 

The more accurate DMC result shows  instead that,
while both DIM and AFI states appear around $\epsilon \approx  10\%$,  
DIM has the lowest  energy 
for all increasing strains until failure.  
Accurate DMC therefore suggests that the charge instability is
dominant over the spin, which is just the opposite of 
what the reference DFT calculation suggested.
In line with that, the prevalence of DIM over AFI is reduced in the less accurate variational Monte Carlo calculations~\cite{SM}.

In addition, the lowest energy will not predict the experimental phase diagram, where isotropic strain $\epsilon$ is obtained 
by tensile stress $\sigma$.  
The equilibrium state under stress, 
rather than energy, minimizes  the enthalpy $H(\sigma) = \min_{S} \left[E(S) - \sigma S \right]$,  
where $ \sigma=\partial_S E(S)$ with $S$ the mean area. 
The stress-area term makes in principle all negative-stress states metastable, 
as an infinitely large enthalpy gain can always be obtained by breaking the lattice apart. 
A metastable stretched state of graphene is, nonetheless, protected against failure by a large barrier, 
connected  with the positive slope of the area-stress curve -- the bulk modulus.
A change of sign of that slope 
signals the vanishing of the barrier, 
ushering in mechanical failure.  

In Fig.~\ref{fig2}(b) the maximum strain is $\epsilon_{\rm max} \sim 15\%$ for the DIM phase, 
actually  close to that obtained in Ref.~\cite{Marianetti} by arbitrarily ignoring spin.  
Interestingly, this stability limit of the DIM phase coincides 
[Fig.~\ref{fig2}(a)] with the prevalence within DFT  of a HEX phase of Fig.~\ref{fig1}(c),   
an artificial  state that foreshadows, as it were, the real mechanical failure in a six-atom cell.  
The structurally undistorted AFI and SEM phases have higher enthalpies and are ruled out at high stress 
[Fig.~\ref{fig2}(b)] despite  their mechanical resilience, until about 20 \% strain.
The QMC-calculated enthalpy of strained graphene, our main result, is 
shown as a function of isotropic tensile stress in Fig.~\ref{fig2}(c). 
Obtained by evaluating  the stress with polynomial interpolation, 
the result is affected by a large 
 statistical error (shaded region), mostly due to the large uncertainty of the stress obtained by 
fitting energy-area curves. With luck however, we reduced  this error by 
means of a rigorous upper bound, which is obeyed by the enthalpy difference of 
any given phase  from the symmetric phase
\begin{equation}
  H(\sigma)-H_{\rm SEM}(\sigma)\leqslant   E(S)-E_{\rm SEM}(S)
  \label{eq.ub}
\end{equation}
where $S$ is the area corresponding to the stress 
$\sigma$ in the symmetric phase. 
The upper bound is practically coincident with the  mean value, 
totally eliminating the error. The 
ground-state 
phase diagram predicted by minimum enthalpy, Fig.~\ref{fig2}(c),  shows that the SEM state for  
 $\sigma < \sigma_{\rm l} = 25.1$ ${\rm N/m}$ 
( $\epsilon_{\rm l} = 8.5\% $)   is followed  by a DIM distorted state for 
$\sigma_{\rm l} < \sigma < \sigma_{\rm u} = 30.4$ ${\rm N/m}$ 
( $\epsilon_{\rm u} = 15\% $) where  stability of the DIM phase is lost, and mechanical failure ensues. 
Even though metastable AFI and SEM phases still persist up to 20\% strain, their realization should imply 
an unphysical enthalpy rise.  One may therefore speculate that the difference between our calculated mechanical failure point, 
and that extracted from indentation ($ \sigma = 40-50$ ${\rm N/m}$,  $\epsilon = 22.5\% $) should be attributed to 
the absence of realistic  indentation details in our total uniform idealized description.

\begin{figure}
  \begin{center}
    \includegraphics[width=0.75\columnwidth]{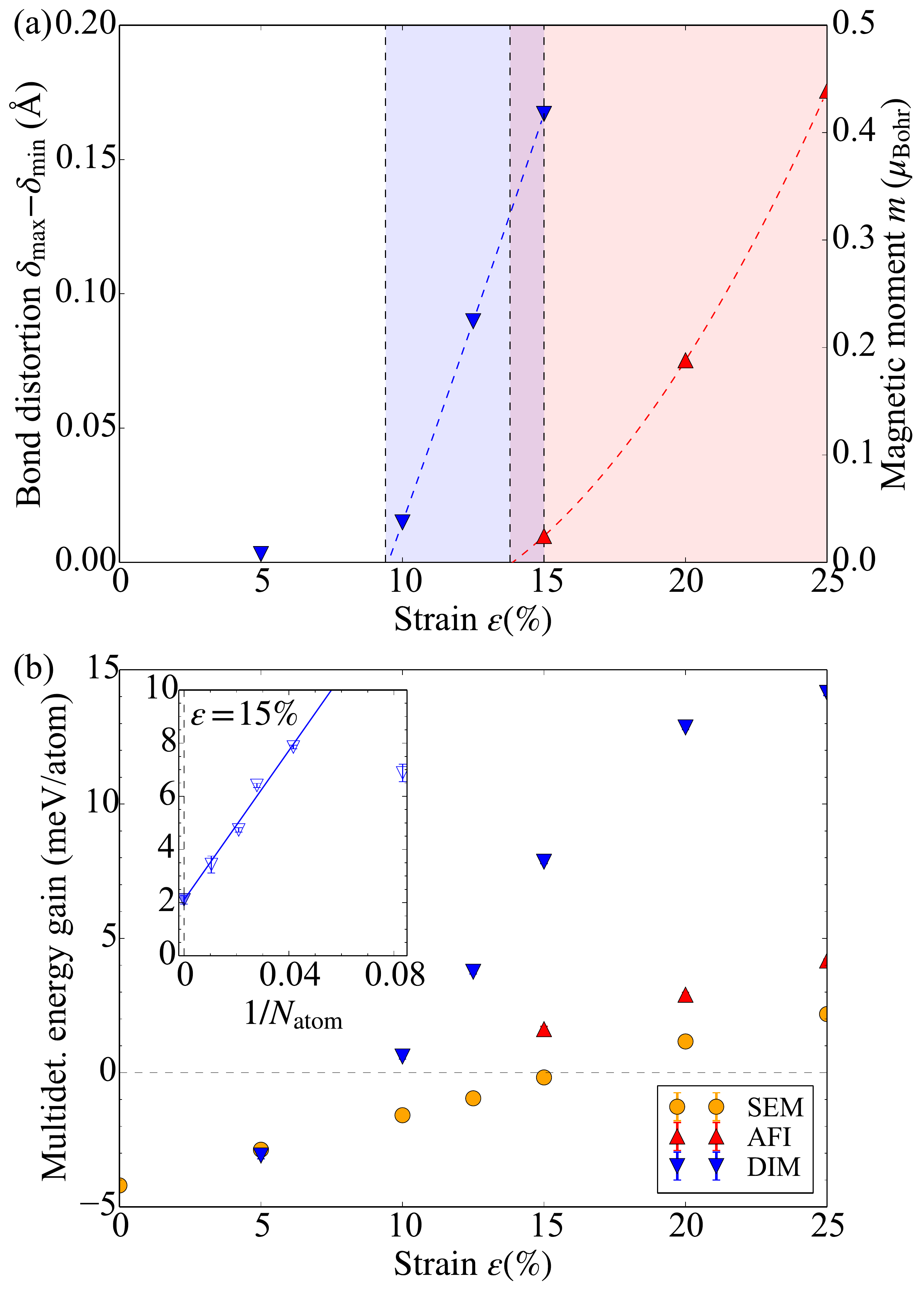}
    \caption{
      \label{fig3}
      (a) Graphene DIM (left axis) and AFI (right axis) order parameters as a function of 
      strain $\epsilon$.       
      The purple shaded area in the vertical lines indicates a DIM-AFI coexistence region. 
      The Heisenberg model limit is $\simeq 0.54 \ \mu_{\rm Bohr}$~\cite{PhysRevB.73.054422}.      
      (b) 
      Correlation energy gain, measured by the energy per atom  difference between the single determinant ansatz 
      (Jastrow-Slater determinant wavefunction) and the corresponding 
      multideterminant  JAGP wavefunction. 
      The largest energy gain occurs in the DIM state, 
      underlining its resonating valence-bond 
      nature, actually increasing for large strain $\epsilon$.
      Small negative values at small strain are finite-size effects.       
      Inset: finite size scaling of this correlation energy gain 
      in the DIM state
      at $\epsilon=15\%$. 
    }
  \end{center}
\end{figure}

We can finally characterize and understand the DIM state, between 8.5 and 15\% strain. 
The dimerizing distortion order parameter of Fig.~\ref{fig3}(a), 
defined as the difference between large and small bond lengths,  
has the Peierls-Kekul\'e symmetry of  Fig.~\ref{fig1}(c) and appears 
to set in continuously, reaching $\sim 0.17$ ${\rm \AA}$ near 
the DIM stability limit $\epsilon_{\rm l} = 15\% $. 

The above QMC results for ground state properties of isotropically stressed graphene raise important physical questions. First, how and why do correlations stabilize the DIM phase instead of the AFI preferred by DFT between $\sim 10$ and 15 \% strain? Second,  what is the electronic gap of
the insulating DIM phase of the strained graphene?  Third, is the DIM insulator topologically trivial or nontrivial and what  consequences does the answer entrain?

To the first point, the multideterminantal character of our variational ansatz of Eq.~(\ref{wf}),  
originally a paradigm for the resonating valence bond (RVB) state~\cite{ Pauling, anderson1973}, 
is crucial for the  enhanced stability of the correlated DIM state. 
The Jastrow factor ${\cal J}$ partly projects out from the determinantal part $\Psi_{\rm AGP}$ 
the single C-C molecular orbital (Mulliken) electron pair  term, which is
largest in unstressed graphene but energetically penalized by electron-electron repulsion under stress. 
That favors the  two-determinant C-C valence bond (Heitler-London) term.   
All goes qualitatively as in the textbook two-electron problem of strained ${\rm H}_2$ molecule.  
A black bond in Fig.~\ref{fig1}(c), with obvious notations, 
is the entangled combination of the two Slater determinants 
$c_{A\uparrow}^\dag c_{B\downarrow}^\dag |0 \rangle$ and  $c_{A\downarrow}^\dag c_{B\uparrow}^\dag |0 \rangle$ 
between $A$ and $B$, with zero double occupancies. 
By contrast, the uncorrelated Peierls molecular orbital 
wavefunction $(c_{A\uparrow}^\dag+c_{B\uparrow}^\dag) (c_{A\downarrow}^\dag+c_{B\downarrow}^\dag)|0\rangle$ 
involves a larger double occupancy for both sites, and a bad 
electron-electron repulsion. This  many-electron entangled wavefunction of the correlated DIM phase 
contains a Jastrow factor acting on an exponentially large number of Slater determinants 
$2^{N_{\rm s}/2}$, that appropriately penalizes the atomic configurations,
where $N_{\rm s}$ is the total number of singlet bonds.  

To gauge the correlation energy gain permitted by our ansatz, we show in Fig.~\ref{fig3}(b) 
the correlation energy obtained  by the multideterminant $\Psi_{\rm AGP}$ over a single determinant, 
still with the Jastrow factor.
This difference is obtained by projecting the pairing function $f$,  
for each twist used, to the optimal $f_P$ obtained by restricting to the best single determinant, 
calculated from the orthogonal eigenfunctions~\cite{marchimol} $\phi_i$ associated to the original pairing function 
$f$  [i.e., $\int d{r'}^3  f({\bf r},{\bf r}') \phi_i({\bf r}')= \lambda_i \phi_i({\bf r})$, where $\lambda_i$ are the corresponding eigenvalues]  
as $f_P({\bf r},{\bf r}') = \sum_{i=1}^{N/2} \lambda_i \phi_i({\bf r}) \phi_i({\bf r}')$  with the largest $|\lambda_i|$~\cite{zenagp}. 
Since $N$ electrons exhaust  the occupation of  the $N/2$ one particle orbitals $\phi_i$, 
$f_P$ describes the corresponding Slater determinant  possessing maximum weight $\prod_i |\lambda_i|$  
in the multideterminant expansion of the AGP, as described in Ref.~\cite{zenagp}. 
The small energy excess of this simpler wavefunction and the full JAGP, computed by correlated sampling, 
measures the multideterminant "RVB" correlation energy gain.   
As shown in Fig.~\ref{fig3}(b), 
this correlation energy gain is negligible in both perfect honeycomb structures, 
i.e., the  poorly strained SEM  and the largely strained AFI phases. 
Conversely,  it becomes 
extensive (see inset) 
and growing with order parameter in the DIM phase, 
which therefore becomes stabilized, rather than the loser as in DFT. 
Stabilization of the DIM phase can be attributed to superexchange energy that is  poorly 
treated within DFT.

To the second point, the electronic gap and the difference between charge and spin gaps 
is not directly obtainable by a QMC ground state calculation, but we get an order of magnitude  from DFT, where the  DIM electronic Kohn-Sham 
gap grows from zero at 8.5\% strain to about 1.1 eV at 15\%~\cite{SM}

\begin{figure}
  \begin{center}
    \includegraphics[width=1.0\columnwidth]{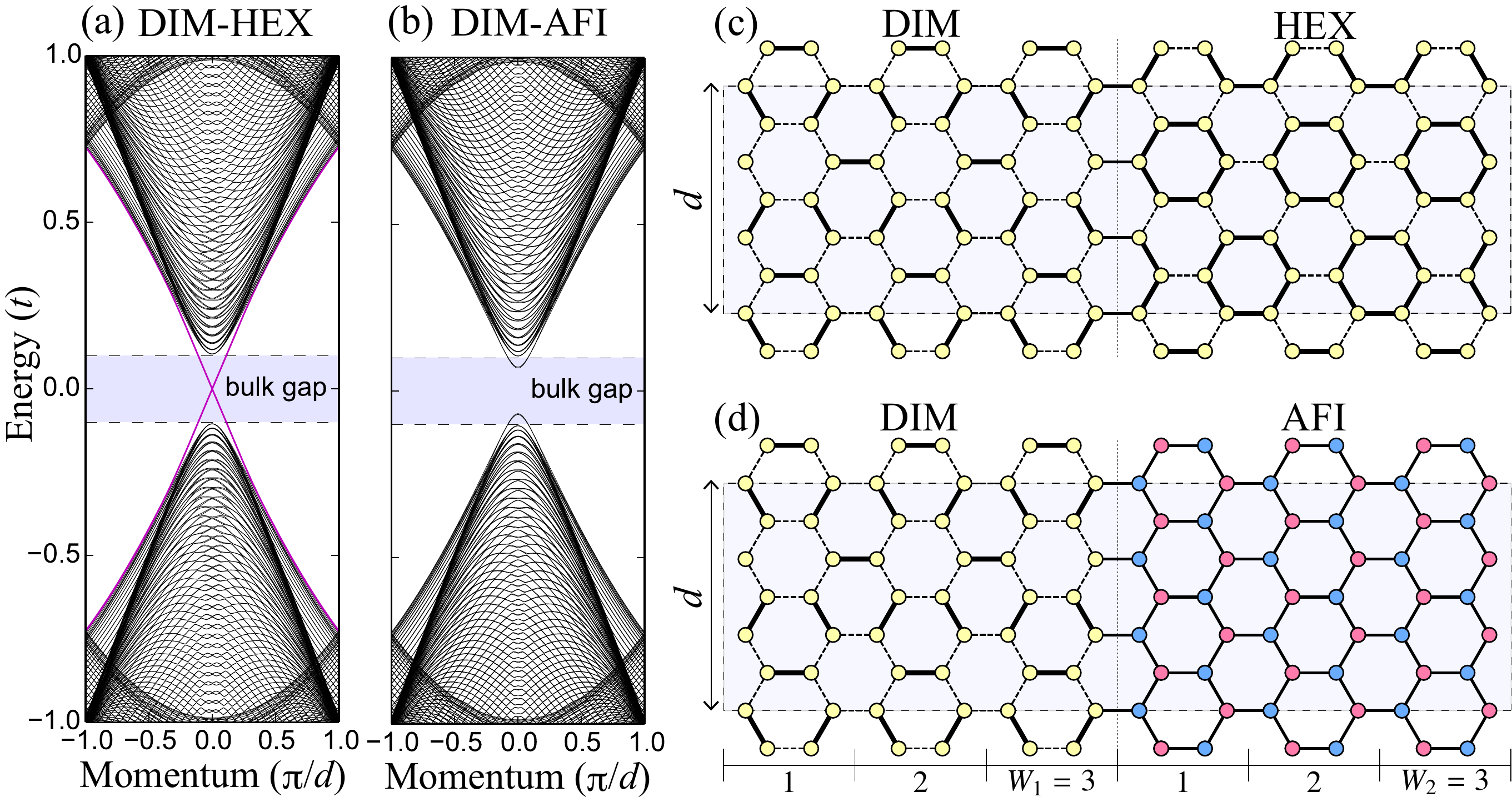}
    \caption{
      \label{fig5}
      Tight-binding band structures of systems with zigzag interfaces between 
      (a) DIM and HEX phases and  
      (b) DIM and AFI phases.   
      The vertical axis is the single-particle excitation energy relative to the Fermi level and 
      the horizontal axis the momentum parallel to the interface. 
      The parameters $(t_A,t_B,t_C)$ are assumed to be 
      $(0.95t,0.95t,1.05t)$ for DIM, $(1.05t,1.05t,0.95t)$ for HEX, and $(t,t,t)$ for AFI. 
      The hopping integrals connecting the two phases are set equal to $t$. 
      The magnitude of the gap in the AFI phase is set equal to that of the DIM phase,  
      as indicated by the shaded area in (a) and (b). 
      The intragap interface modes are highlighted with thick magenta lines in (a).              
      Schematic figures of the calculated interface between 
      (c) the DIM and HEX phases and 
      (d) the DIM and AFI phases.       
      The dotted-vertical line in (c) and (d) indicates the zigzag interface, 
      and the shaded area the unit cell. The size of the unit cell is determined by the 
      widths $W_1$ and $W_2$ of the two phases, indicated in the bottom of (d), and 
      the vertical length $d$. 
      The calculations in (a) and (b) are for $W_1=W_2=50$. 
      The system as a whole lies on a torus.
    }
  \end{center}
\end{figure}

To the third point, we note that adiabatic continuity between the strongly correlated DIM state and 
the uncorrelated Kekul\'e state discussed in literature~\cite{Kariyado2017, Frank2011, Liu2017} implies 
the DIM insulating state of strained graphene is topologically nontrivial, unlike the AFI or HEX  states. 
With reference to Fig.~\ref{fig1} (one-electron tight-binding is sufficient for this purpose), 
the bond dimerization of the DIM phase corresponds to 
$|t_C| > |t_A| = |t_B|$, 
while the HEX phase to 
$|t_C| < |t_A| = |t_B|$. 
The nontrivial nature of the DIM phase is protected by 
the sublattice (chiral) symmetry and the mirror symmetry along a bond~\cite{Kariyado2017}. 
While this fact has no special consequences in infinite perfect 2D  graphene strained into the 
DIM phase, it will, as in other topological insulators~\cite{reviewtopo}, show up at 
interfaces and defects, which can support a topological state 
energetically placed inside the dimerization gap. 
As a demonstration of that, we present a model tight-binding 
DIM-HEX two-phase coexistence with the zigzag interface under 
periodic-boundary conditions 
[Fig.~\ref{fig5}(c)]. 
Its  electronic structure  in Fig.~\ref{fig5}(a) shows topological states, 
with their characteristic gapless modes crossing the Fermi level, 
localized at the two DIM-HEX interfaces. This is in contrast to a model DIM-AFI interface [Fig.~\ref{fig5}(d)] 
where no gapless interface states appear [Fig.~\ref{fig5}(b)]. 
This difference is simply understood because the DIM-HEX system preserves 
the two symmetries described above but the DIM-AFI system does not. 
Although the bulk single-particle gap increases with the electron correlations, 
these topological features remain qualitatively the same because of  
adiabatic continuity.

The impact of increasing electron correlations in isotropically and uniformly strained graphene, 
calculated by QMC simulations
with an accurate variational wavefunction, is in summary predicted to be nontrivial.  
The phase diagram dictated by minimizing enthalpy under increasing stress predicts the sequence: 
SEM-DIM-failure, different from the best spin-polarized density-functional  predictions.  
Large electron correlations stabilize the DIM phase, schematized in Fig.~\ref{fig1}(c), 
in the 8.5-15\% tensile strain range corresponding to 25-31${\rm N/m}$ stress range. 
Roughly speaking, dimerization freezes Pauling's resonating valence bond, 
a state
which fluctuates  
in the honeycomb spin-liquid state as described, e.g., by Ref.~\cite{Meng2010}, 
into a valence-bond solid, realized by a Kekul\'e-like phase that breaks translation invariance. 
Remarkably this effect was very recently observed in a lattice model of bilayer graphene~\cite{palee_2018}.  
The DIM phase possesses a stress-dependent order parameter and a correspondingly increasing electronic gap. 
In correspondence with the predicted continuous SEM-DIM transition the mechanical impedance of graphene should exhibit a dissipation singularity.
Electronically,  the graphene DIM insulator is topological, implying protected
intragap states localized around   defects with peculiar symmetry properties, 
including topological 1D Dirac states at grain boundaries and dislocations.  
Our predicted 15\% failure strain is somewhat smaller than the 22.5\% reported 
by experimental indentation studies, possibly due to the role of nonuniformities in 
indentation mechanics,  absent in our so far totally uniform calculations. 
The onset of the DIM structural deformation and of an electronic gap which DFT estimates 
in the order of about one eV at failure, as well as of topologically related defect states 
in this gap could be used in the future to detect spectroscopically this novel state of strained graphene.

\acknowledgments
The authors thank T. Neupert for helpful discussions. 
Computational resources were provided by K computer at RIKEN Center for Computational Science (R-CCS) 
through the HPCI System Research project
(Projects No.~hp160126, No. hp170079, and No. hp170308) 
and PRACE-13-3322. 
S.S. acknowledges the kind hospitality at RIKEN R-CCS, in the early stage of this work. 
K.S. acknowledges support from the JSPS Overseas Research Fellowships. 
E.T. acknowledges support by ERC Advanced Grant No. 320796 - MODPHYSFRICT. 
T.S. ackwnoledges the Simons Foundation for funding. 

\bibliography{biball}

\clearpage
\clearpage
\setcounter{page}{1}
\newcommand{\beginsupplement}{%
        \setcounter{table}{0}
        \renewcommand{\thetable}{S\arabic{table}}%
        \setcounter{figure}{0}
        \renewcommand{\thefigure}{S\arabic{figure}}%
}

\beginsupplement

\onecolumngrid
\begin{center}
  {\large{\textbf{
        Supplemental Material for\\
        Correlation-Driven
        Dimerization and  Topological Gap Opening 
        in Isotropically Strained Graphene
  }}}  
  
  \vspace{0.2 cm}
  Sandro Sorella$^{1,2,3}$, 
  Kazuhiro Seki$^{1,3,4}$, 
  Oleg O. Brovko$^{5}$,
  Tomonori Shirakawa$^{1,3,4,6}$, \\
  Shohei Miyakoshi$^{6}$,
  Seiji Yunoki$^{3,4,6}$, and
  Erio Tosatti$^{1,2,5}$ 
  
  \vspace{0.2 cm}
{\small
  {\it
    $^1$ {International School for Advanced Studies (SISSA), via Bonomea 265, 34136, Trieste, Italy} \\
    $^2$ {Democritos Simulation Center CNR--IOM Istituto Officina dei Materiali, Via Bonomea 265, 34136 Trieste, Italy} \\
    $^3$ {Computational Materials Science Research Team, RIKEN Center for Computational Science (R-CCS),  Hyogo 650-0047,  Japan} \\
    $^4$ {Computational Condensed Matter Physics Laboratory, RIKEN Cluster for Pioneering Research (CPR), Saitama 351-0198, Japan} \\
    $^5$ {The Abdus Salam International Centre for Theoretical Physics (ICTP), Strada Costiera 11, 34151, Trieste, Italy} \\
    $^6$ {Computational Quantum Matter Research Team, RIKEN, Center for Emergent Matter Science (CEMS), Saitama 351-0198, Japan} 
  }
}
\vspace{0.2 cm}
\end{center}

*{Density-functional-theory calculation}
First principles calculations were carried out within density-functional theory (DFT) 
based on the projector-augmented-wave method~\cite{Blochl1994} and a plane-wave 
basis set~\cite{Kresse1996} as implemented in the Vienna Ab-initio Simulation Package (VASP)~\cite{Kresse1993,Kresse1996}. 
Exchange and correlation were treated with the hybrid HSE functional~\cite{Heyd2003} 
known to perform well for carbon materials and even for small gap semiconductors~\cite{Paier2006}. 
For bulk and lattice constant calculations an energy cutoff of $600~\mathrm{eV}$ for the plane wave expansion and 
a Monkhorst-Pack $k$-point mesh~\cite{mp_grid} with $21\!\times\!21\!\times\!1$ points (before symmetry operations 
application) were used. For both uniform expansion and dimerized configurations a six atoms orthorombic supercell was used. 
In the calculations of the dimerized and hexagonal phases all atoms were allowed to relax until 
the residual forces were smaller than $10^{-3}~\mathrm{eV/\AA}$. 
For electronic convergence an energy variation criterion was uses with a threshold of $10^{-7}~\mathrm{eV}$.

\section*{Quantum Monte Carlo calculation}
%{\color{red}  
The Jastrow-AGP ansatz used in this work [Eq.~(1) in the main text] was indeed proposed by P. W. Anderson to describe the Pauling's RVB\cite{anderson1973}, in particular the  benzene molecule. This is very simple to understand by modeling this molecule with a 6-site Heisenberg model on a ring, with a spin 1/2 $\pi-$electron on each site of the lattice. In this case the $6 \pi-$electrons wavefunction, in second quantized form, reads:
\begin{equation}
\Psi_{\rm JAGP}= P_G \exp\left(\sum\limits_{i,j=1}^6  f_{i,j} c^\dag_{i \uparrow} c^{\dag}_{j\downarrow} \right) |0 \rangle
\end{equation}
where $c^\dag_{i,\sigma}$ creates an electron with spin $\sigma$ on the lattice sites $i=1,\cdots 6$ and $P_G$ is the Gutzwiller projection, a simplified 
version of the Jastrow correlation, that in this case projects out 
configurations with two electrons with opposite spins in the same site.
By taking a pairing function $f_{i,i+1}=1=f_{i+1,i}$ and otherwise zero, 
the rotational symmetry of the model is not broken by the ansatz.
It is then simple to realize that, after applying the Gutzwiller projection 
to this particular pairing function $f$, we obtain exactly the two Kekule' valence bonds with the same amplitude, as a general property of the so called short-range RVB (nearest-neighbors atoms singlet pairing) described in details in Ref.~\cite{yunoki}.
  On the other hand by breaking the symmetry of the pairing function, one can favor one of the
 two structures, that in the Heisenberg model is not energetically favorable. In the realistic calculation of benzene the same effect was found in the first application of the
 Jastrow AGP wavefunction~\cite{casulaagp}
where the RVB energy was found lower than the broken symmetry pairing function, corresponding to a dominant Kekule structure.
%}

In the realistic calculation of graphene, 
we expand the two pairing functions $f({\bf r},{\bf r'})$ and $u({\bf r},{\bf r'})$ 
defining the variational wavefunction $\Psi_{ \rm JAGP}$, 
over a finite localized basis set 
(5s3p1d for $f$ and 3s2p for $u$), 
and  minimize the total energy by the simultaneous optimization of the Jastrow factor and
the determinantal part, within a consistent stochastic approach~\cite{ourwork}. 
The Jastrow factor is initialized by the uncorrelated limit ${\cal J}=1$, 
whereas the initial determinantal part $\Psi_{\rm AGP}$ is obtained 
by using the DFT with the local-density or 
the local-spin-density approximations (LSDA). 
The initial trial atomic positions are generated by scaling the 
undistorted equilibrium ones by a fixed factor 
ranging from $1$ to $1.25$.
Standard pseudopotentials~\cite{bukowski} are used to remove 1s core electrons, 
as they do not affect the chemical bond. 
%%%%
Moreover, when AFI phase is studied, the pairing function is no longer 
symmetric~\cite{groswf}, 
as in this way we can describe also the unrestricted Slater determinant of an LSDA calculation, 
that we use to initialize the optimization of our ansatz.
%%%%

In most calculations we consider 24 carbon atoms in an orthorombic 
supercell with rectangular basis, whereas for the size scaling of Fig.3b (main text) we use also 12, 36, 48, and 96 carbon supercells. 
In order to minimize finite-size effects, we adopt the 
twist-averaged-boundary conditions~\cite{Lin2001,Dagrada2016,Karakuzu2017} in the $x$ and $y$ directions with averaging over a $6\times 8$ Monkhorst-Pack grid~\cite{mp_grid}, 
whereas in the $z$ direction, we adopt simple periodic-boundary conditions with a very large distance 
($300$ Bohrs) between the graphene images. 
%%%
In order to minimize the number of variational parameters and achieve 
a faster and smoother convergence in the thermodynamic limit,  
the pairing function 
is parametrized with a given set of variational parameters, with the same 
values for all different twists,  as discussed in Ref.~\cite{arxiv}.  
Obviously this choice provides an higher energy than optimizing independently 
each twist with different variational parameters, but this residual gain should obviously vanish in the thermodynamic limit, because a given twist cannon change the energy in an extensive way.
However this effect can explain why in a finite size calculation, the multideterminant energy gain shown in Fig.3b of the main text, a non negative  quantity in the thermodynamic limit,
could be slightly negative.
%%%

We verified that,with this setup, in the 24 carbon supercell, 
the graphene structure is 
in exact agreement with experiments, with a lattice constant of $a_0=1.414$ ${\rm \AA}$, and a perfectly isotropic honeycomb lattice, despite the  rotational symmetry breaking boundary conditions.
The number of variational parameters is reduced by exploiting translation symmetry for this system. In our 24 atom supercell three cases are possible: i) the standard unit cell with two identical atoms; ii) the same unit cell, now with AFM polarization of the two atoms; iii) a larger unit cell with six atoms, compatible with all allowed lattice distortions predicted by the Frank-Lieb 
theorem~\cite{Frank2011}. 
In all cases, we first optimize energy, by relaxing all variational 
parameters defining the Jastrow factor and the determinantal part, 
together with  the atomic positions within the 
constant-volume (and shape) supercell simulation. 
The optimization of the atomic positions is done  
with an efficient method based on the covariance-matrix of the nuclear forces, 
which allows us to determine their equilibrium positions efficiently and accurately~\cite{accelerated}. 
We also employ lattice-regularized diffusion Monte Carlo (DMC) within the fixed-node approximation, using a lattice mesh of $a_{\rm mesh} = 0.2,0.3,$ and $0.4$, respectively, and extrapolated the results  for $a_{\rm mesh} \to  0$ in the standard way. The fixed-node approximation is necessary for fermions for obtaining statistically meaningful ground-state properties. In this case the correlation functions/order parameters, depending only on local (i.e., diagonal in the basis) operators, such as the ones presented in this work, are computed with the forward walking technique~\cite{forward}, which allows the computation of pure expectation values on the fixed-node ground state.

\section*{Thermodynamic phase diagram of graphene under tensile strain with variational Monte Carlo}
\begin{figure}[!ht]
  \begin{center}
    \includegraphics[width=0.5\columnwidth]{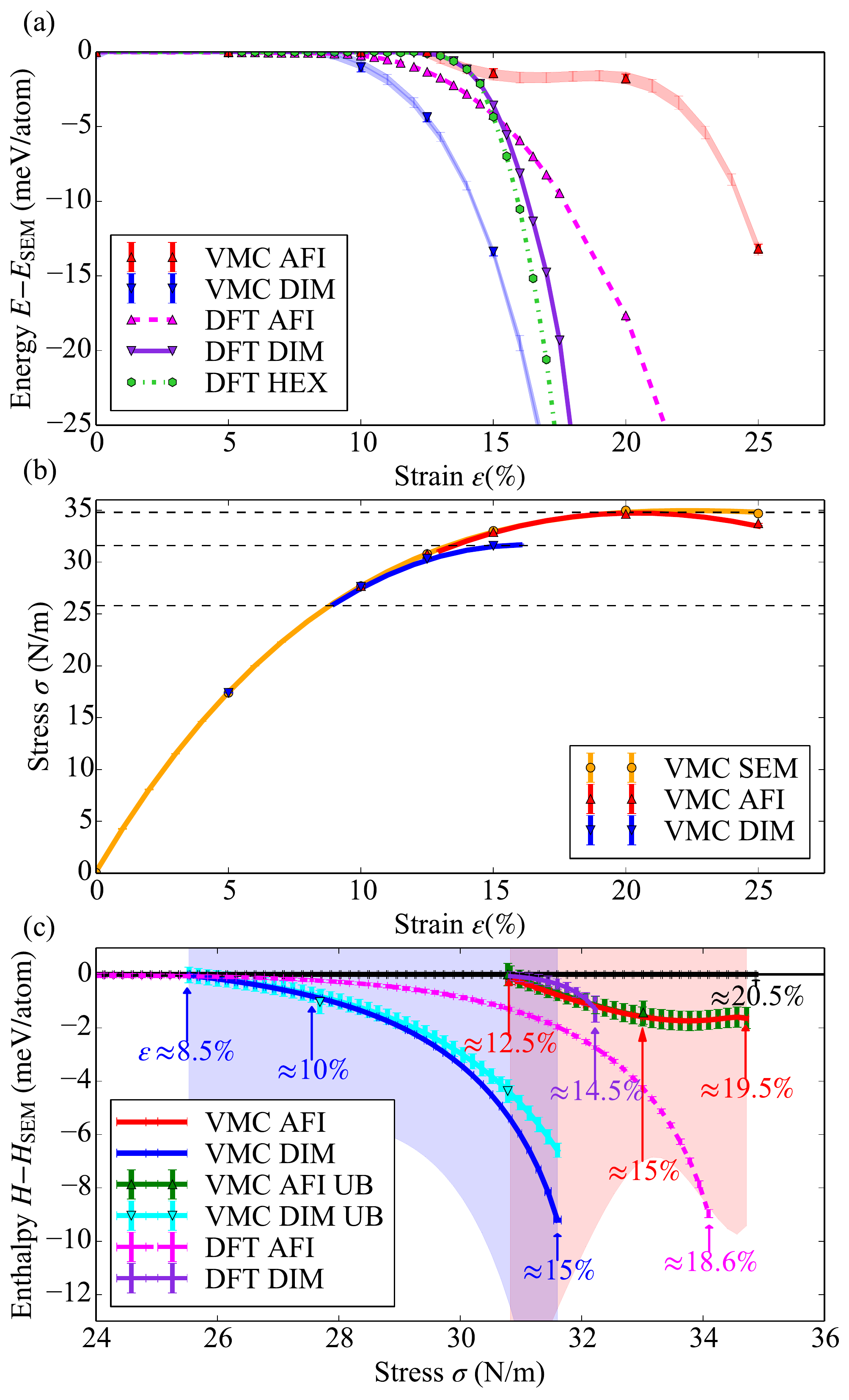}
    \caption{
      \label{fig_s1}
      (a) Total energy $E$ relative to the SEM phase $E_{\rm SEM}$
      as a function of strain $\epsilon$ obtained by DFT and by VMC 
      for the DIM, AFI, and HEX phases.
      (b) Stress ($\sigma$)-strain ($\epsilon$) curve for graphene obtained by VMC.
      Dashed lines  mark the transition stress values $\sigma_{\rm l}$ and $\sigma_{\rm u}$ 
      for SEM-DIM (continuous)  DIM-AFI (discontinuous), AFI-failure.
      (c) The enthalpy $H$ relative to the SEM phase $H_{\rm SEM}$ as a function of tensile stress $\sigma$.
      The blue- and red-shaded regions indicate the error bars on the enthalpies for DIM and AFI phases by VMC.
      The upper bounds of the relative enthalpy for the DIM and AFI phases are also shown (DIM UB and AFI UB) 
      with much smaller error bars. 
      The corresponding strain $\epsilon$ for the several selected points (indicated by arrows) are also shown. 
    }
  \end{center}
\end{figure}

Figure~\ref{fig_s1} shows the results of the thermodynamic phase diagram 
by the variational Monte Carlo (VMC) method. 
As shown in Fig.~\ref{fig_s1}c, 
the VMC shows that the AFI phase is stabilized for $\sigma \gtrsim 31.7 {\rm N/m}$, while 
the DMC does not, as shown in Fig.~2 of the main text.  
This implies that the AFI phase is very unlikely because with a better accuracy the AFI phase is less stable. 
Moreover, the enthalpy gain of the DIM phase by the DMC is more enhanced than the VMC. 
All these findings consistently indicate  that the main aspects of our phase diagram should be essentially robust against further improvements in the description of the correlation energy. 

Close to $\epsilon=15\%$, there may be some very tiny region with coexistency between DIM and AFI phases. 
However, the energy difference between the two phases is smaller than our resolution, limited by the statistical errors.
Also, the small order parameter obtained at $\epsilon = 5\%$ for the DIM state 
could be removed by a more accurate optimization, 
that we cannot afford with QMC, as too many iterations with high statistical accuracy are required.

\section*{Single-particle gap}
\begin{figure}[!ht]
  \begin{center}
    \includegraphics[width=0.8\columnwidth]{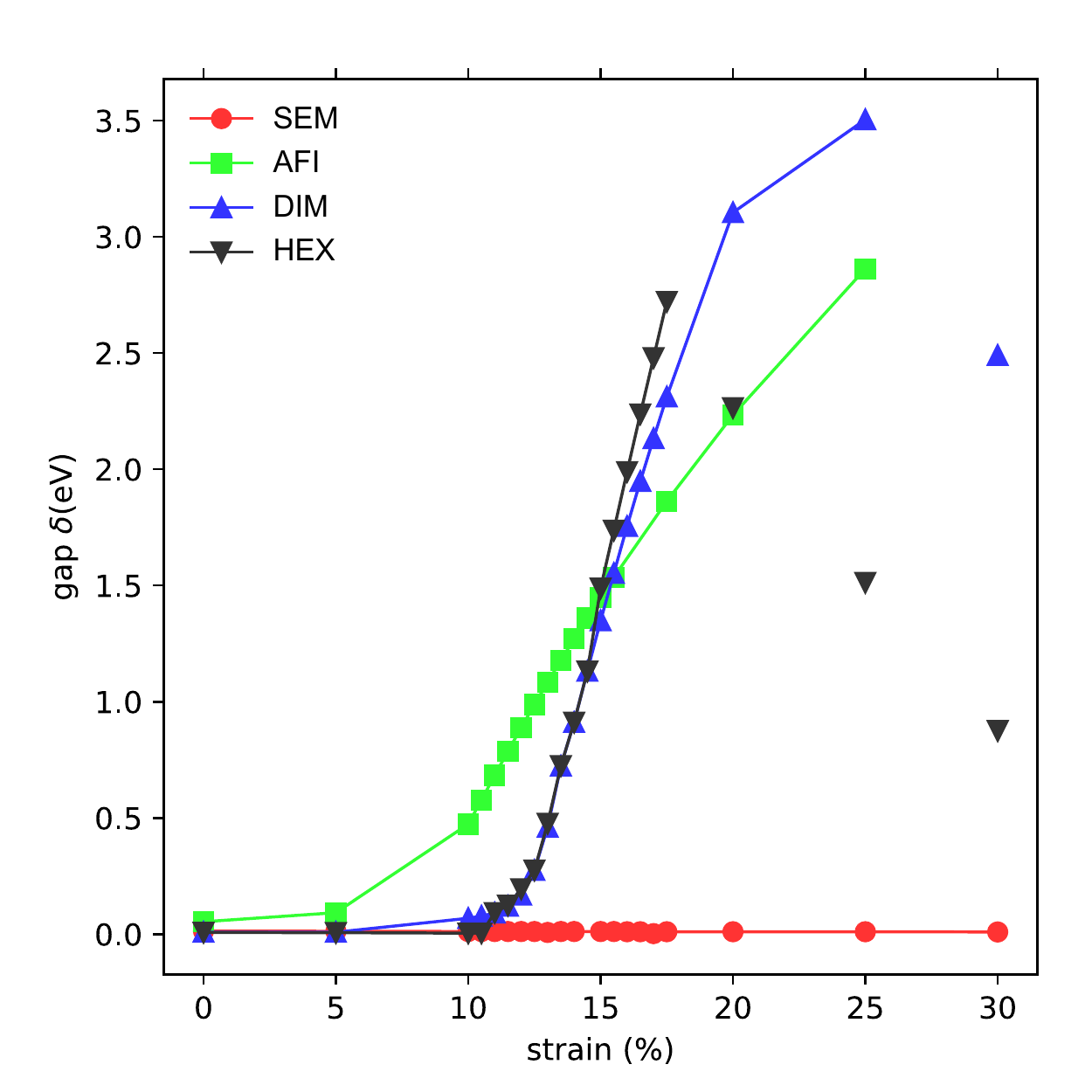}
    \caption{
      \label{fig_s2}
      The single-particle gap as a function of the strain 
      for SEM (red circles), AFI (green squares), DIM (blue triangles), and HEX (black-inverted triangles) phases 
      obtained by DFT. 
    }
  \end{center}
\end{figure}

Figure~\ref{fig_s2} shows the single-particle gap as a function of the strain 
for SEM, AFI, DIM, and HEX phases obtained by DFT.

\section*{Bond length}
\begin{figure}[!ht]
  \begin{center}
    \includegraphics[width=0.8\columnwidth]{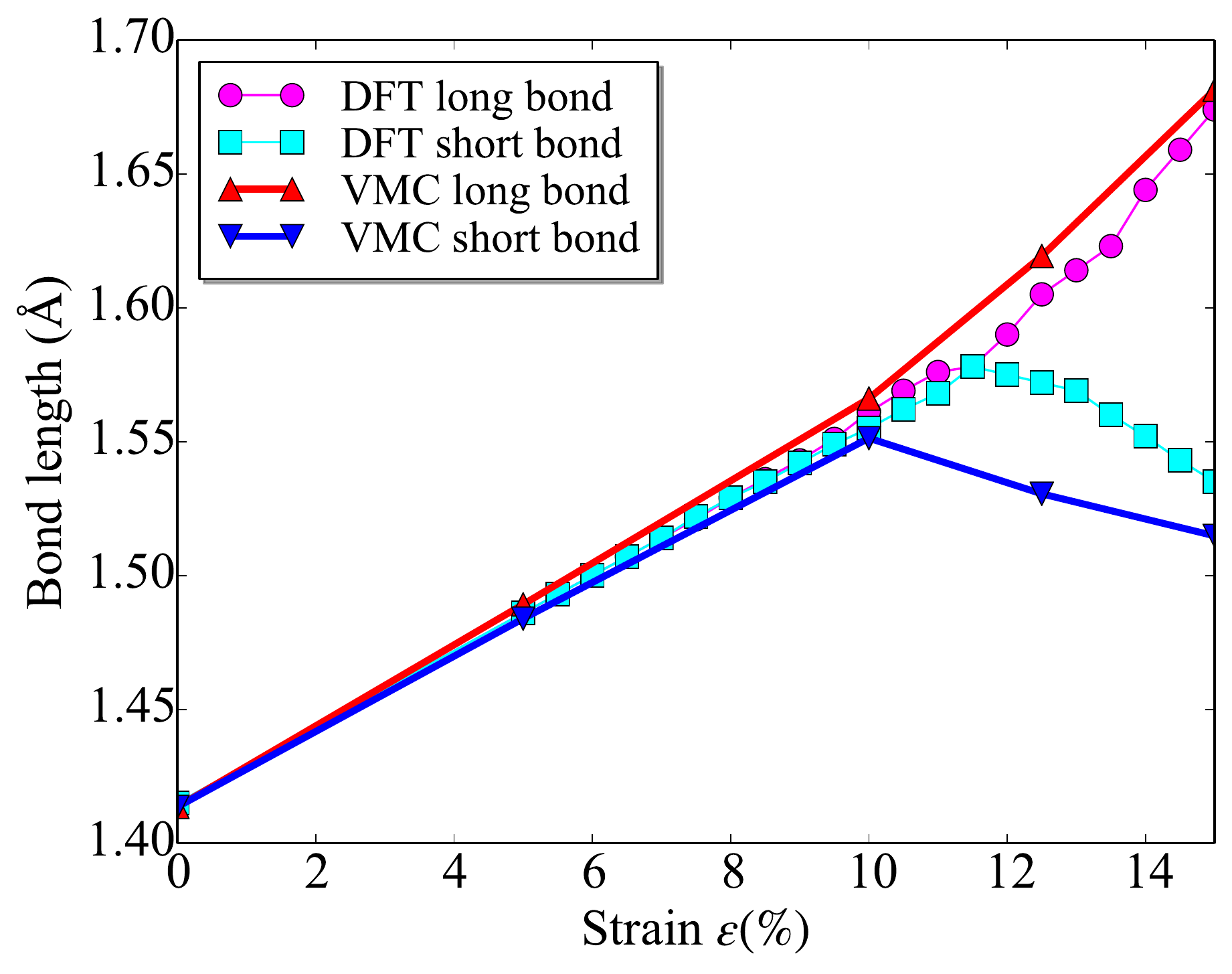}
    \caption{
      \label{fig_s3}
      The bond lengths for DIM phase obtaind by VMC (regular and inverted triangles) and 
      DFT (circles and squares). 
    }
  \end{center}
\end{figure}

Figure~\ref{fig_s3} shows the carbon-carbon distance on the short and long bonds (bond length) 
for DIM phase obtained by VMC and DFT. 

\end{document}